\documentclass[reprint,superscriptaddress,showpacs,amssymb,floatfix]{revtex4-1}
\usepackage[final]{graphicx}
\usepackage{amsmath}
\usepackage{siunitx}

\begin{document} 

\title{Super-directional light emission and emission reversal from micro cavity arrays}

\author{Jakob Kreismann}
\email{jakob.kreismann@tu-ilmenau.de}
\affiliation{Institute for Physics, Theoretical Physics II/Computational Physics Group, Technische Universit\"{a}t Ilmenau, Weimarer Stra\ss{}e 25, 98693 Ilmenau, Germany}
\author{Jaewon Kim}
\affiliation{Institute for Physics, Theoretical Physics II/Computational Physics Group, Technische Universit\"{a}t Ilmenau, Weimarer Stra\ss{}e 25, 98693 Ilmenau, Germany}
\author{Mart\'{\i} Bosch}
\affiliation{Institute for Physics, Theoretical Physics II/Computational Physics Group, Technische Universit\"{a}t Ilmenau, Weimarer Stra\ss{}e 25, 98693 Ilmenau, Germany}
\author{Matthias Hein}
\affiliation{Department of EI, Electrical Engineering and Information Technology, RF and Microwave Research Group,Technische Universit\"{a}t Ilmenau, Helmholtzplatz 2, 98693 Ilmenau, Germany}
\author{Stefan Sinzinger}
\affiliation{Department of Mechanical Engineering, Optical Engineering Group, Technische Universit\"{a}t Ilmenau, Helmholtzring 1, 98693 Ilmenau, Germany}
\author{Martina Hentschel}
\affiliation{Institute for Physics, Theoretical Physics II/Computational Physics Group, Technische Universit\"{a}t Ilmenau, Weimarer Stra\ss{}e 25, 98693 Ilmenau, Germany}
\date{\today}

\begin{abstract}
\textbf{Optical microdisk cavities with certain asymmetric shapes are known to possess unidirectional far-field emission properties. Here, we investigate arrays of these dielectric microresonators with respect to their emission properties resulting from the coherent behaviour of the coupled constituents. This approach is inspired by electronic mesoscopic physics where the additional interference effects are known to enhance the properties of the individual system. As an example we study the linear arrangement of nominally identical Lima\c{c}on-shaped cavities and find mostly an increase of the portion of directional emitted light while its angular spread is largely diminished from 20 degrees for the single cavity to about 3 degrees for a linear array of 10 Lima\c{c}on resonators, in fair agreement with a simple array model. Moreover, by varying the inter-cavity distance we observe windows of reversion of the emission directionality and super-directionality that can be interesting for applications. We introduce a generalized array factor model that takes the coupling into account. 
}
\end{abstract}

\pacs{
42.55.Sa, 
42.60.Da, 
}

\maketitle

\textbf{Introduction.} 
Optical microdisk cavities have become a target of intense studies because of their versatile properties to control light, and their unique application potential \cite{Vahala2003} as high-Q resonators \cite{Michael2007}, microlasers \cite{McCall1992}, \cite{Harayama2011} or sensors\cite{Armani2007}, \cite{Yang2018}. Moreover, they play an important role as widely accessible experimental realizations of theoretical model systems~\cite{Cao2008} used in quantum chaos~\cite{Stockmann1999,Stone1997,An2010,Wiersig2011} and mesoscopic physics~\cite{Datta1995}.  

In view of the mutual inspiration of nonlinear dynamics, quantum chaos, antenna theory and the field of optical microcavities (a paradigm example being the intimate relation between the unstable manifold of the cavity and its far-field emission characteristics~\cite{Lee2005,Wiersig2008}), it is suggestive to take further inspiration from mesoscopic physics and its tool box. Besides the wide field of the manipulation of geometric phases~\cite{Berry1984,M.Nakahara1990} in dielectric systems~\cite{Bliokh2008,Ma2016,Kreismann2018}, extension of the single system into array or ensemble structures is known to enhance specific properties in electronic mesoscopic systems. For example, side structures of Aharonov-Bohm peaks in magnetoconductance oscillations were resolved in arrays of electronic micro-rings due to self-averaging in the composite system~\cite{Nitta1999}. 

In the present work, we want to adopt the concept of arrays to optical microcavities. Our naive expectation and motivation is the possibility of an enhancement of directional emission properties inherent to single deformed optical microdisk cavities, e.g. of the Lima\c{c}on shape~\cite{Wiersig2008}. 
To this end, we investigate a linear array (chain) of Lima\c{c}on resonators. The shape of each Lima\c{c}on resonator is given in two-dimensional polar coordinates $(r,\phi)$ by $r(\phi)=R\cdot(1+\delta \cos(\phi) ) $,
where $R$ and $\delta$ represent the mean radius and the deformation parameter, respectively. The resonator itself is assumed to be purely dielectric with a high refractive index of $n=3.0$ embedded in vacuum, $n_0=1$. The deformation parameter is set to $\delta=0.43$, a value known~\cite{Wiersig2008} to provide highly directional far fields.

We use MEEP (MIT Electromagnetic Equation Propagation~\cite{MEEP}), a free finite-difference time-domain (FDTD) software package for electromagnetic wave simulations, to calculate the normalized resonance frequencies $\Omega=kR=\text{Re}(\omega)R/c$ of the Lima\c{c}on resonator, with $\omega$ being the complex frequency and $c$ the speed of light, the quality factor $Q=-0.5\text{Re}(\omega)/\text{Im}(\omega)$, the distributions of the electric field component $E_{z}(x,y)$ (modes) and the far-field intensity $I(\theta)$ with far-field angle $\theta$. In the present work, we study wavelength-scale cavities. For this reason, $kR\sim7$, with $k = 2 \pi / \lambda$ being the wave number and $\lambda$ the wavelength in vacuum. We use $E_{z}$-point-dipole sources (electric field $E$ perpendicular to the cavity) placed off-center to excite the same mode in each cavity and focus on TM-polarized modes.
In addition we use COMSOL~\cite{comsol} in order to calculate the eigenfrequency  splitting of the resonator array, which results from the exchange of energy among adjacent resonators.\\
We indeed observe the expected enhancement of the far-field emission of Lima\c{c}on arrays in comparison to the single resonator due to the coherent operation of all microcavities in the array. We describe this effect in the first part of the paper for the weak (evanescent) coupling regime. 
However, under certain conditions we observe the opposite behaviour, i.e., a reversal of the main emission direction, or super-directional emission, as a result of the collective action. We investigate this strong-coupling regime in the second part of the paper and conclude with a summary.\\
We point out that the excitation scheme used in the MEEP simulations is such that 
all resonators are excited equally (symmetric configuration, \textquotedblleft even mode\textquotedblright) close to the eigenfrequency of the single cavity. Since the excitation pulse in each resonator has a finite spectral width that is slightly broader than the spectral splitting of the array modes for the not too small (and experimentally relevant) $d/R$-values considered here, we observe an excitation close to the symmetric array mode with intensity distributions close to that of the single cavity, cf. Supplemental Material SEC. I~\cite{SM}, and dedicate the following discussion to this case.

\begin{figure}
	\centering
	\includegraphics[width=8.5cm]{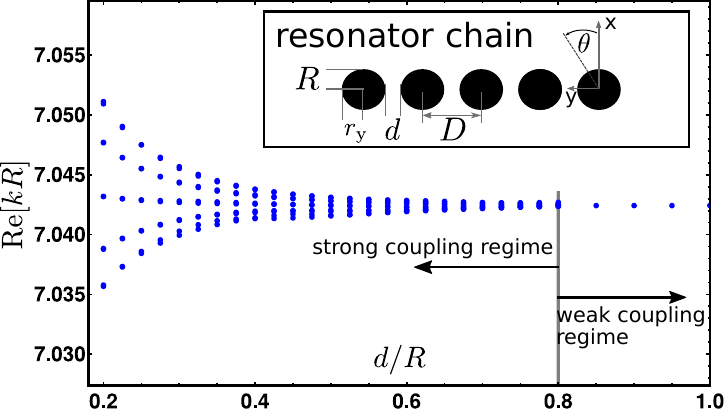}
	\caption{Frequency splitting of a linear 5-element Lima\c{c}on resonator array as function of the normalized inter-resonator distance $d/R$ calculated by COMSOL~\cite{comsol}. $D(d)= 2r_{\text{y}}+d$. Two different coupling regimes can be clearly distinguished. The strong-coupling ($d/R<0.8$) regime exhibits 
	five pairs of eigenfrequencies 
	corresponding to five symmetric and five anti-symmetric array modes. The pairs are hardly  visible, and appear as almost single dots.
	}
	\label{fig:f1}
\end{figure}

{\bf Array emission in the weak-coupling regime.}
We first investigate a row of Lima\c{c}on resonators with large inter-resonator distance 
$d/R\sim1$. In this regime, 
the eigenfrequencies do not split, see FIG.~\ref{fig:f1}, indicating that the resonators do not exchange energy. The entire array of resonators can thus be treated as a superposition of individual resonators:

\begin{equation}
	\Phi_{\text{array}}=\sum_{n=1}^{N}\alpha_{n} \psi_{n}(x,y-y_n),
	\label{equ:chain1}
\end{equation}

where $\Phi_{\text{array}}$ describes the electric (or magnetic) field distribution of the resonator array, and $\psi_{n}(x,y-y_n)$ are the electric (or magnetic) 
eigenmodes of a single Lima\c{c}on resonator shifted along the y-axis by $y_n=(n-1)D$. The complex-valued coefficients $\alpha_{n}$ represent the amplitude and phase of the $n$-th resonator. In order to compute the far fields (fields in the Fraunhofer region) of the array, we use the near-field-to-far-field-transformation, a common method in FDTD~\cite{FDTD-Inan2011}. 
From Eq.~(\ref{equ:chain1}) follows the z-component $E_{\text{z,array}}^{\text{FF}}$ of the electric far field (see Supplemental Material SEC. II~\cite{SM}): 

\begin{equation}
	E_{\text{z,array}}^{\text{FF}} (\theta) \propto E_{\text{z,single}}^{\text{FF}}(\theta) \sum_{n=1}^{N}\alpha_{n}e^{-ikD(n-1)\sin{\theta}},
	\label{equ:AFmodel}
\end{equation}

where $E_{\text{z,single}}^{\text{FF}}$ is the z-component of the electric far field of a single resonator, and $\theta$ represents the far-field angle, see FIG~\ref{fig:f1}. The sum on the right-hand side of the equation is also known as the array factor $AF(\theta)$ from antenna theory~\cite{AntennaTheory}. The array factor describes the complex-valued far-field pattern of a uniform linear array of point sources. Thus, the total far field results from that of a single resonator multiplied by the array factor. Indeed, our configuration is very similar to antenna arrays. 
%
%
\begin{figure}
	\centering
	\includegraphics[width=8.5cm]{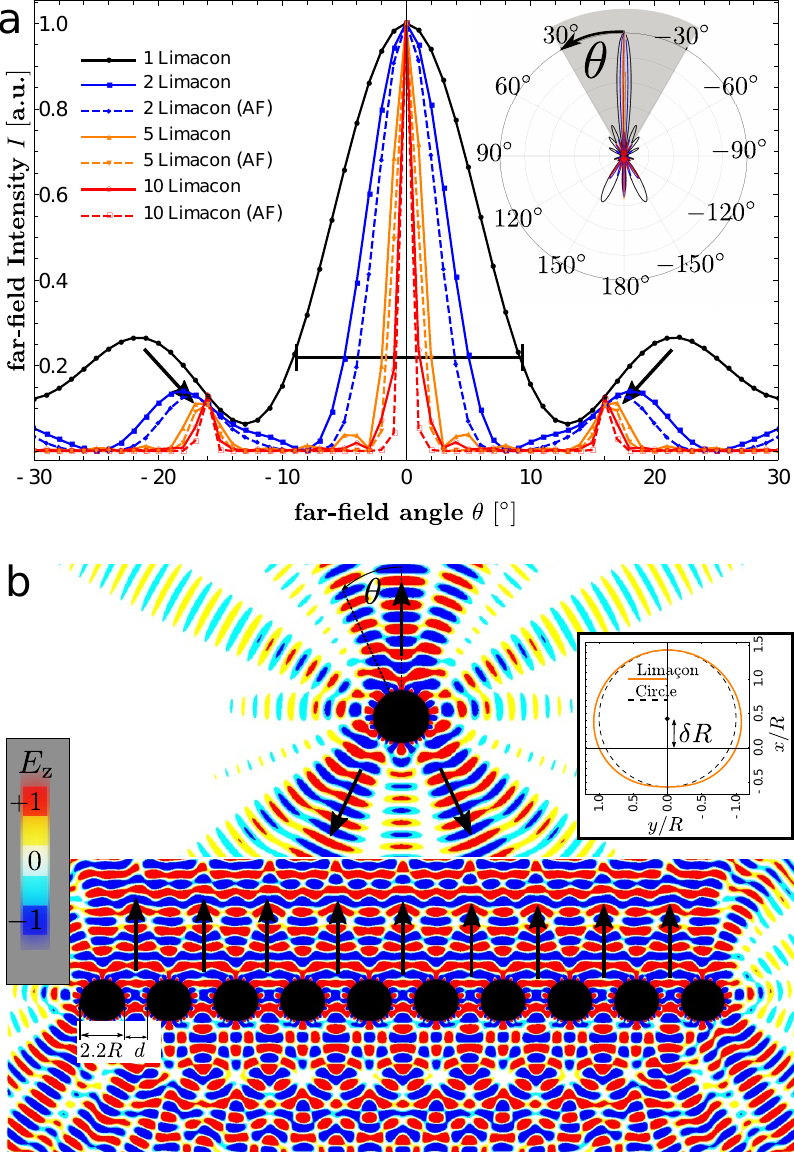}
	\caption{\textbf{a:} Intensity plot of the far-field emission from Lima\c{c}on cavity arrays. The intensity plot displays the far-field main lobe within the range highlighted by the gray triangular region in the inset (polar plot of the far-field intensity). \textbf{b:} Field distribution of the $E_\text{z}$ component of the electric field of 1 and 10 Lima\c{c}on resonators, respectively. The inter-resonator distance is $d/R=1$. The inset shows the Lima\c{c}on-shape and for comparison a circle with radius $R$ centered at $x=\delta R$.
	}
	\label{fig:f2}
\end{figure}
First, we wish to study the difference between the far fields of the single resonator and of an array. 
Figure~\ref{fig:f2} a shows the far-field intensity $I\propto\left|E_{\text{z,array}}^{\text{FF}} (\theta)\right|^2$ depending on the far-field angle $\theta$ within the range of $\pm 30$ degrees for 1, 2, 5 and 10 resonators aligned in a row with constant inter-resonator distance $d/R=1$. The solid lines are the results from the full FDTD calculation whereas the dashed lines represent results assuming the array factor model according to Eq.~(\ref{equ:AFmodel}). 
The inset presents the full polar plot of the far-field intensity. The gray region marks the $\pm 30$ degrees range. 

The agreement between these results suggests that 
the resonator array behaves as expected from an antenna array in the weak-coupling regime considered here \footnote{The coefficients $\alpha_{n}$ have been all set to unity because there is no mutual coupling between the resonators, and all resonators have been excited equally.}.
As a result, the width of the main lobe decreases with increasing number of array elements (resonators) yielding increased directivity. This feature conforms quantitatively to the array factor model. 
Specifically, we observe that the width of the main lobe of the far-field emission from 10 Lima\c{c}on resonators is reduced by roughly one order of magnitude compared to the single resonator case. 
Furthermore, we observe that the position and height of the side lobes varies with the number of elements, saturation at $N > 5$ (cf.~the black arrows). This behaviour results from the array factor as well  (see Supplemental Material FIG. 2~\cite{SM}~for more details).

Figure~\ref{fig:f2} b displays the normalized $E_\text{z}$-field distribution 
for one and 10 Lima\c{c}on resonators, respectively.
The far field of one Lima\c{c}on resonator (upper panel) is well known and very similar to the results e.g.~in~\cite{Song2011}. The lower panel shows the situation for
10 resonators and reveals a strong emission in forward direction as indicated by the black arrows. This enhancement arises from the superposition of individual, unperturbed resonator modes of the single Lima\c{c}on cavity within the array.\\
\begin{figure}
	\centering
	\includegraphics[width=8.5cm]{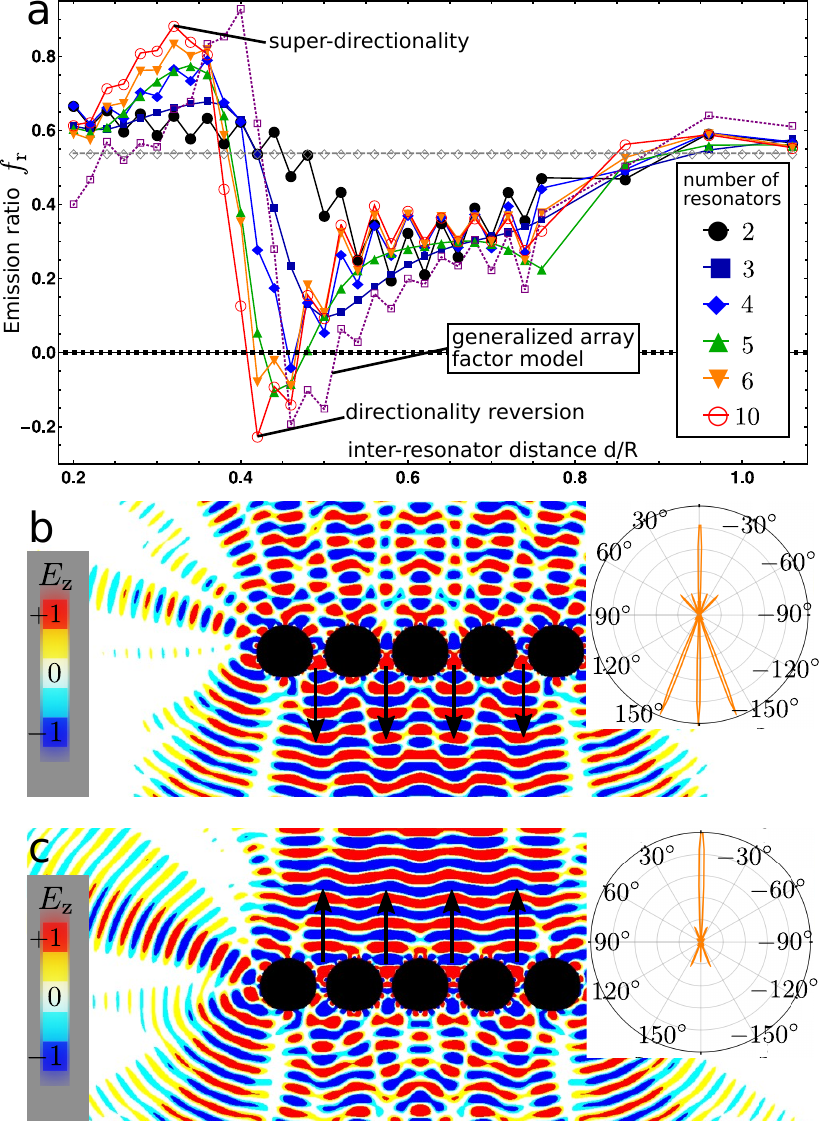}
	\caption{\textbf{a:} Emission ratio $f_{\text{r}}$ as function of the inter-resonator 
	distance ratio $d/R$ for arrays with 2,3,4,5,6, and 10 resonators. \textbf{b and c:} spatial distribution of the normalized electric field amplitude $E_\text{z}$  
	for the 5-resonator array outside the cavities and for distance values (b) $d/R = 0.44$ (minimum of $f_{\text{r}}$ -- reversal of directionality) and (c) $d/R= 0.34$ (maximum of $f_{\text{r}}$ -- super-directional light emission in forward direction).  
	Insets show the polar far-field emission. See text for details. 
	}
	\label{fig:f3}
\end{figure}

{\bf Array emission in the strong-coupling regime.}
We now investigate the emission of the resonator array for intercavity distances $d/R$ smaller than 1, i.e.~encountering (strong) element coupling. Here, 
the resonators are close enough to exchange energy by mutual coupling. As a result, the frequencies of the single resonators split, and array modes emerge, see FIG.~\ref{fig:f1} and Supplemental Material FIG. 1~\cite{SM}. 
Naturally, the array factor model, Eq.~(\ref{equ:chain1}, \ref{equ:AFmodel}), remains no longer valid. Due to the coupling, the properties of coupled cavities differ from the isolated case.
However, a generalization of this model (generalized array factor model) takes the coupling of the resonators into account by replacing $\psi_n$ and $E_{\text{z,array}}^{\text{FF}}$ of the bare single cavity with the respective quantities $\psi_n^\text{c}$ and $E_{\text{z,array}}^{\text{FF,c}}$ of the single coupled cavity. The far field of the resonator array in the strong coupling regime is modeled by multiplying the far fields of the single coupled cavity with the array factor:
\begin{equation}
E_{\text{z,array}}^{\text{FF,c}} (\theta) \propto E_{\text{z}}^{\text{FF,c}}(\theta) \sum_{n=1}^{N}\alpha_{n}e^{-ikD(n-1)\sin{\theta}}.
\label{equ:GenArray}
\end{equation}   

\noindent The generalized array factor model can qualitatively predict the observed array far fields with all features described below while emphasizing the role of coupling.

In order to quantify the directional emission of a resonator array, we introduce the far-field emission ratio $f_{\text{r}}$ that is defined by:
\begin{equation}
f_{\text{r}}:=\frac{f_+ - f_-}{f_+ + f_-},
\label{equ.EmissionRatio}
\end{equation}
where 
$f_+=\int_{-10^\circ}^{10^\circ}I(\theta)d\theta$ is the forward directional emission and $f_-=\int_{170^\circ}^{190^\circ}I(\theta)d\theta$ represents the backward directional emission. 
Emission ratios $f_{\text{r}}$ greater (smaller) than $0$ 
represent forward (backward) directionality, whereas values around $0$ indicate balanced forward and backward emission. By means of the ratio $f_{\text{r}}$, we study the far-field behaviour of the resonator array as a function of the inter-resonator distance $d/R$ which is displayed in Figure~\ref{fig:f3} a. 

{\bf Reversal of emission directionality.} 
According to FIG.~\ref{fig:f3} a we observe that decreasing $d/R$ 
and hence increasing the coupling strength, reduces the emission ratio $f_{\text{r}}$, i.e., less light is emitted into the forward direction. For arrays consisting of more than three resonators, we find the emission ratio $f_{\text{r}}$ to change sign, indicating backward emission and thus a reversal of the emission directionality. For the mode chosen and $N=10$ resonators, we find $f_{\text{r}}$ to reach a minimum of around $f_{\text{r,min}}\approx-0.22$ at a cavity spacing of $d/R \approx0.42$.\\ 
The purple dashed line shows the result of a 10-resonator array obtained with the generalized array factor model ($\psi_n^\text{c}$ has been taken from a 4-resonator array) which nicely agrees with the results from the full FDTD calculation. \\
In order to investigate this behavior in more detail, we look at the spatial distribution of the $E_\text{z}$-component of the electric field amplitude as depicted in FIG.~\ref{fig:f3} b. 
We clearly see a wave propagating downwards below the resonator array (resembling a plane wave propagation) as indicated by the black arrows. Above the resonator row, we observe wave propagation in form of an interference pattern that indicates partially destructive interference. The inset displays the entire far-field emission as a polar plot, and confirms that more energy is emitted backwards than forwards.\\
Assuming coupling between the resonators to be the reason for the directionality reversion, then an interesting question is: Why do 5 resonators show reversion, but 2 resonators do not? 
Following the $f_{\text{r}}$-curve of 2 resonators in FIG.~\ref{fig:f3} a, we observe that the emission ratio decreases a bit with decreasing $d/R$-values but its minimum remains positive. This slight decrease of $f_{\text{r}}$ indicates an increase of the energy emitted into the backwards direction. It turns out that directionality reversion is already present for 2 resonators but it is masked by the emission from the edges of the array (see Supplemental Material FIG. 4~\cite{SM}). The directionality reversion arises from the coupling regions (in-between space of two resonators) and it needs a certain number of these coupling regions to overcome the effect of the emission from the edge regions.\\
{\bf Super-directional light emission from linear arrays.} 
By decreasing the distance $d/R$ between the resonators further, the emission ratio changes sign and reaches a maximum $f_{\text{r,max}}\approx0.89$ at $d/R=0.32$ for $N=10$ resonators. This maximum is even higher than the value found at the element distance  $d/R\approx1.0$. It indicates that the coherent modal action of the resonator array results in a super-directional forward emission of light. This is illustrated in Figure~\ref{fig:f3} c.
Contrary to Figure~\ref{fig:f3} b, we now find (almost) planar wave propagation into the forward direction. The interference pattern visible below the resonator row indicates destructive interference and hence reduced energy flow into the backward direction, as confirmed by the inset showing the full polar plot. \\
We point out that the emission patterns in Figs.~\ref{fig:f2} b and \ref{fig:f3} c look similar as both correspond to values $f_{\text{r}}>0$. Notice however that the planar-wave-type propagation (destructive interference) in forward (backward) direction are even more pronounced for the strong coupling case shown in Fig.~\ref{fig:f3} c. 

\begin{figure}
	\centering
	\includegraphics[width=8.5cm]{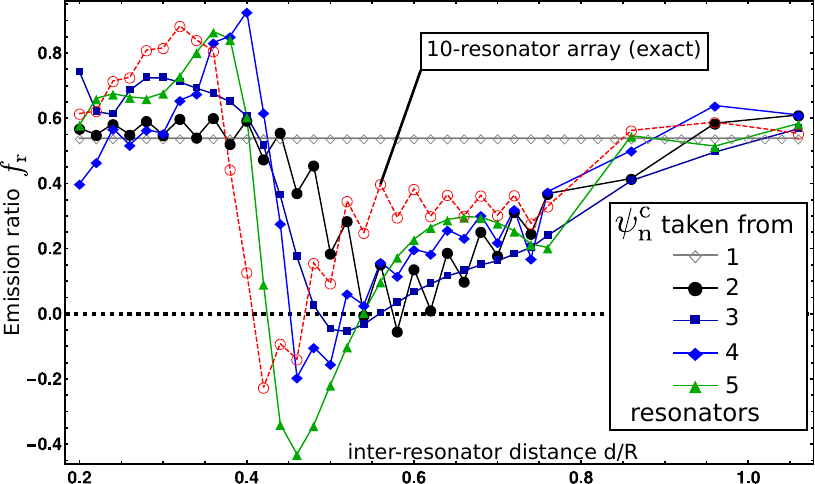}
	\caption{ Results of the emission ratio $f_{\text{r}}$ calculated from the generalized array factor model for $N=10$, cf. Equ.~(\ref{equ:GenArray}), where $\psi_n^\text{c}$ was taken from a 1, 2, 3, 4, and 5 element resonator array, respectively. The dashed red line represents the \textquotedblleft exact\textquotedblright~emission ratio of a 10-element resonator array. 
	}
	\label{fig:f4}
\end{figure}

The gray dashed line in FIG.~\ref{fig:f3} a corresponds to the (standard) array factor model (Equ.~\ref{equ:AFmodel}) with $N=10$ resonators. It represents the case of weakly coupled resonators and shows a constant $f_\text{r}$-value because of $\left|AF(\theta)\right|^2=\left|AF(\pi+\theta)\right|^2$ and $E_{\text{z,single}}^{\text{FF}}(\theta)$ being independent from $d/R$. \\
{\bf Generalized array factor model.} 
We now study the generalized array factor model in more detail. In this context, an interesting question is: Is it possible to predict the features of a 10-resonator array with the aid of the generalized array factor model utilizing the coupling features of a 2 or 3-resonator array?\\
To this end, we compute the emission ratio for a 10-resonator array based on the far fields obtained from the generalized array factor model where the $\psi_n^\text{c}$ have been taken 
from a 1, 2, 3, 4, and 5 element resonator array, respectively (see Supplemental Material FIG. 3~\cite{SM}). The corresponding results are shown in FIG.~\ref{fig:f4} with a comparison to the exact emission ratio of a 10-resonator array. The case of $\psi_n^\text{c}$ taken from 1 resonator is identical to the array factor model of uncoupled resonators, as described by Equ.~(\ref{equ:AFmodel}). \\
\noindent The emission ratio with $\psi_n^\text{c}$ taken from 2 resonators displays very weak directionality reversion and very  weak super-directionality. At least, one could guess that such features exist. As pointed out earlier, the directionality reversion feature is already present for 2 resonators but it is masked by the emission from the edge regions.  
The best agreement with the exact curve is obtained for the case with $\psi_n^\text{c}$ taken from 4 resonators.
Interestingly, the results with $\psi_n^\text{c}$ taken from 5 resonators
show too much directionality reversion compared to the exact result. The reason for this is that we neglected the role of the amplitudes $\alpha_{n}$. For reasons of simplicity, the $\alpha_{n}$ have been set to unity which is not correct if the array gets larger, because of the amplitude distribution within the array mode with high amplitudes for central resonators and low amplitudes for edge resonators.\\
\noindent The generalized array factor model can qualitatively predict the emission ratio of a 10-resonator array and reveals that the reversion of the directionality arises from the coupling of the single resonator modes. It cannot be modeled by a proper superposition of the single resonator modes (as described in equation~\ref{equ:AFmodel}). Thus, the coupling between the resonators plays an essential role. This implies immediately that the reversed and super-directional emission will be resonance dependent, see Supplemental Material FIG. 5~\cite{SM}. A complementary interpretation of our findings based on (generalized) phase-space arguments~\cite{Wiersig2008} is in progress and will be subject of a subsequent study.\\

\textbf{Conclusion.} 
Our FDTD calculations revealed that the directionality of the emission of photonic devices can be considerably enhanced using array structures, both concerning the fraction of the directionally emitted light and its angular spread. Moreover, the main emission direction of the cavity ensemble can be inverted such that light is mostly emitted into the direction opposite the one of the singly cavity. We argue that this effect arises from the coupling of the individual resonators. The directionality properties depend strongly on (i) the normalized distance $d/R$ between the resonators and (ii) the properties of the individual resonant modes. Furthermore, the directionality reversion behavior depends on the number of the resonators in the array, and emerges only if their number is larger than three. \\

\begin{acknowledgments}
	This work was partly supported by Emmy-Noether programme of the German Research Foundation (DFG). Furhtermore, we thank Christoph Wagner from the Department of Advanced Electromagnetics  for fruitful discussions.
\end{acknowledgments}


%

\end{document}


\begin{center}
{\Large{\textbf{Supplemental Material }}} \\ 
for \textquotedblleft Super-directional light emission and emission reversal from micro cavity arrays\textquotedblright	
\end{center}

\noindent \textbf{Contents:}\\
~\ref{sec:1}. Array modes and excitation scheme used in MEEP \\
~\ref{sec:2}. From far fields to array factors\\
~\ref{sec:3}. Generalized array factor model\\
~\ref{sec:4}. Directionality reversion depending on number of array elements\\
~\ref{sec:5}. Resonance dependence of directionality reversion and super-directionality

\section{Array modes and excitation scheme used in MEEP}
\label{sec:1}

\begin{figure}
	\centering
	\includegraphics[width=8.5cm]{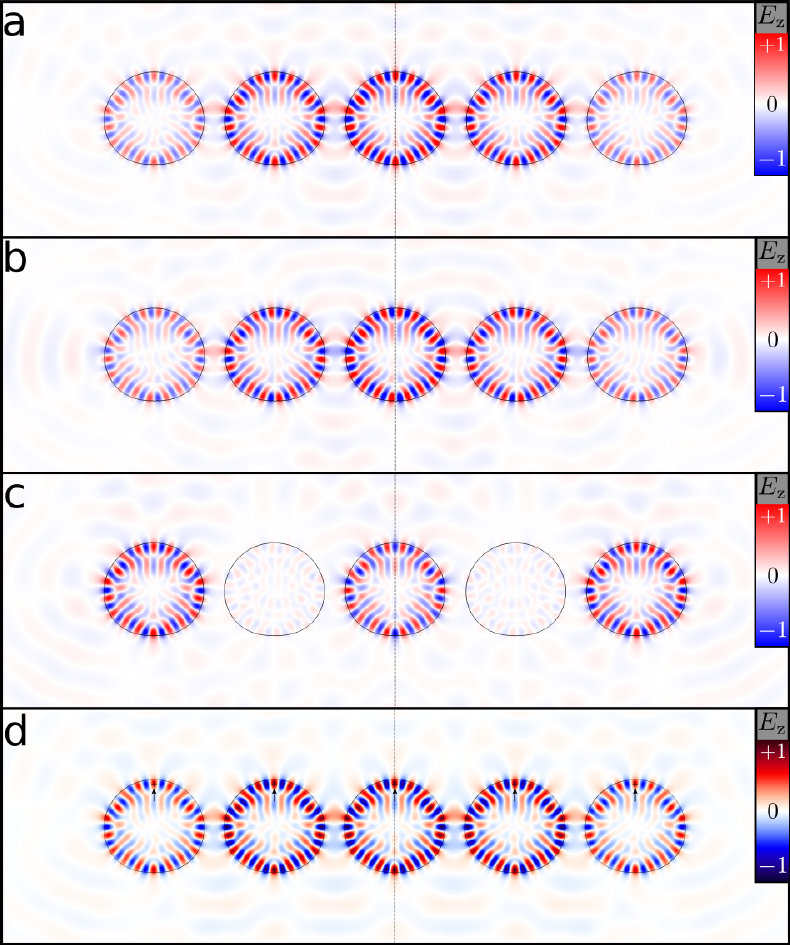}
	\caption{\textbf{a - c:} Field distributions $E_{\text{z}}$ of TM-Eigenmodes of a resonator array with inter-resonator distance $d/R=0.45$ calculated by COMSOL. \textbf{a:} symmetric array mode (w.r.t. the dashed vertical line) with symmetric mode pattern in each resonator. \textbf{b:} anti-symmetric array mode with anti-symmetric mode pattern in each resonator. \textbf{c:} symmetric array mode with outermost resonators being in opposite phase w.r.t the central resonator. \textbf{d:} state after exciting each resonator at the single resonator eigenfrequency calculated in MEEP, see text for further details. 
	}
	\label{fig:S1}
\end{figure}

\noindent The strong-coupling regime exhibits five pairs of eigenfrequencies corresponding to five symmetric and five anti-symmetric array modes. Three examples of TM-polarized array modes calculated by COMSOL are displayed in FIG.~\ref{fig:S1} a - c. The array modes shown in FIG.~\ref{fig:S1} a and b look like joining together symmetric and anti-symmetric single resonator modes, whereas the mode pattern displayed in FIG.~\ref{fig:S1} c exhibits two resonators with almost no field inside. 
FIG.~\ref{fig:S1} d shows the field distribution after equally exciting each resonator at the single resonator Eigenfrequency calculated in MEEP.\\ 
In order to excite the resonators, a point source was placed inside each resonator at the positions indicated by black crosses and arrows in FIG.~\ref{fig:S1} d. Each point source was emitting a TM-polarized ($E_\text{z}$-field) Gaussian pulse centered at the single resonator Eigenfrequency $f_{0,\text{meep}}=1.12056$ with a pulse width $\Delta f_{\text{meep}}=0.01$. After the Gaussian pulse decayed, the simulation was running for additional 100 periods. Following this, the fields were extracted and plotted, and the far fields were computed. The relation between the MEEP-frequencies and the normalized frequency $kR$ used in the manuscript is given by $kR=2\pi f_\text{meep} R_{\text{meep}}$ whereas $R_{\text{meep}}$ is the value of $R$ used in the MEEP-simulation. Usually, this value is set to unity.

\newpage
\section{From far fields to array factors}
\label{sec:2}

\noindent In order to compute far fields, we use the so called near-field-to-far-field transformation, a common method in FDTD software. Exemplarily, we want to compute the far field of a TM mode. The corresponding transformation for the z-component of the electric field in a far distance $r$ is:
\begin{equation}
	E_{\text{z}}^{\text{FF}}(\theta) = 
	 A \oint_{\text{C}} \left(\omega\mu_0 \textbf{e}_\text{z}\cdot \textbf{J}_\text{eq}(\textbf{r}')- k [\textbf{e}_\text{z}\times \textbf{K}_\text{eq}(\textbf{r}')]\cdot\textbf{e}_\text{r} \right) \cdot 
	 \exp\left(-i k \textbf{e}_\text{r}\cdot \textbf{r}'    \right) ds',
\end{equation} 

\noindent where $A$, $\mu_0$, $\omega$ and $k$ is a complex $r$-dependent factor, the magnetic field constant, the frequency and the wave number, respectively.  $\textbf{e}_\text{z}$, $\textbf{e}_\text{r}$, $\textbf{J}_\text{eq}$ and $\textbf{K}_\text{eq}$ represent the unit vector in z-direction, the unit vector into the $r$-direction spanning the far-field angle $\theta$ with respect the x-axis, the equivalent electric and the equivalent magnetic currents, respectively. $\textbf{r}'$ is the position vector of the fictitious boundary of the near fields (position of equivalent currents). The integration is performed along a closed contour surrounding, e.g. the scattering or emitting object.\\
Now, let us consider a resonator array in the weak-coupling regime. Hence, the far field of the array 
is a superposition of the far fields of each resonator: 

\begin{equation}
E_{\text{z,array}}^{\text{FF}}(\theta) = 
\sum_{n=1}^{N} \alpha_n \oint_{\text{C}_n} f_n(\textbf{J}_{\text{eq},n},\textbf{M}_{\text{eq},n})  \exp\left(-i k \textbf{e}_\text{r}\cdot \textbf{r}'_n    \right) ds'_n,
\end{equation} 

\noindent where each resonator is shifted by a distance $D$ (distance between the centers of two resonators). Thus, $\textbf{r}'_n = \textbf{r}'+ (n-1) D \textbf{e}_\text{y}  $ if the resonators are shifted along the y-direction, and $ds'_n = ds' $. Since we are exciting the same mode in each resonator, all the near fields have to be the same. As a result the sum simplifies as follows:

\begin{equation}
E_{\text{z,array}}^{\text{FF}}(\theta) = 
 \sum_{n=1}^{N} \alpha_n \oint_{\text{C}} f(\textbf{J}_{\text{eq}},\textbf{M}_{\text{eq}})  \exp\left(-i k \textbf{e}_\text{r}\cdot \textbf{r}'    \right) ds' \cdot
  \exp\left(-i k D (n-1) \textbf{e}_\text{r}\cdot\textbf{e}_\text{y}     \right).
\end{equation} 

\noindent The integration inside the sum represents the far field of a single resonator placed at the origin and may be evaluated outside the sum. After rearranging the remaining terms, we obtain:

\begin{equation*}
E_{\text{z,array}}^{\text{FF}} (\theta) = E_{\text{z,single}}^{\text{FF}}(\theta) \sum_{n=1}^{N}\alpha_{n}e^{-ikD(n-1)\sin{\theta}}.
\end{equation*}

\begin{figure}
	\centering
	\includegraphics[width=8.5cm]{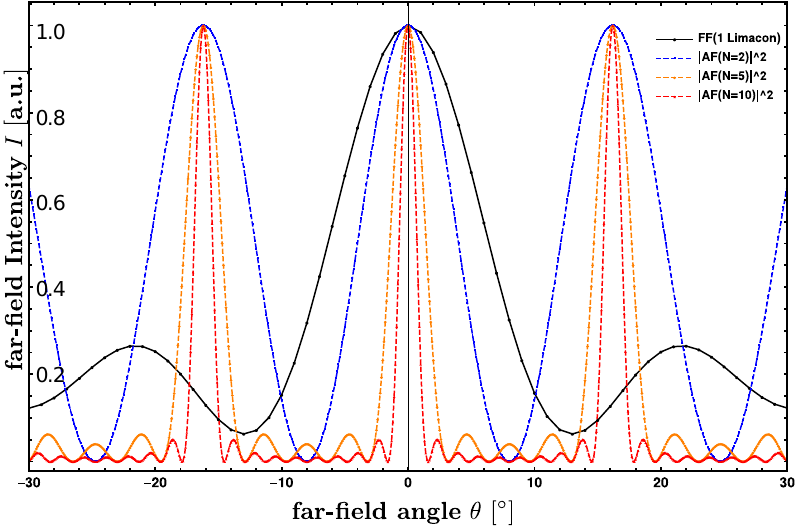}
	\caption{Comparing the far field intensity $I\propto\left|E_{\text{z,single}}^{\text{FF}}(\theta)\right|^2$ and the square of the absolute values of the array factors $\left|\sum_{n=1}^{N}\alpha_{n}e^{-ikD(n-1)\sin{\theta}}\right|^2$ for $N=2,5,10$. Please note, the far field of the single Lima\c{c}on and the squared absolute values of the array factors have been normalized to 1.
	}
	\label{fig:S2}
\end{figure}

\noindent The sum on the right-hand side represents the so called array factor. FIG.~\ref{fig:S2} shows the squared absolute values of the array factor for 2,5 and 10 resonators, as well as the far-field intensity of a single resonator. We see that the width of the peaks of the array factors is reduced clearly from 2 to 5, but is changing slightly from 5 to 10. This slight change of the width is the reason why the side peaks of the array far field seems to saturate as displayed in FIG. 2 of the manuscript.

\vspace{1cm}
\section{Generalized array factor model}
\label{sec:3}

\begin{figure}[h]
	\centering
	\includegraphics[width=11cm]{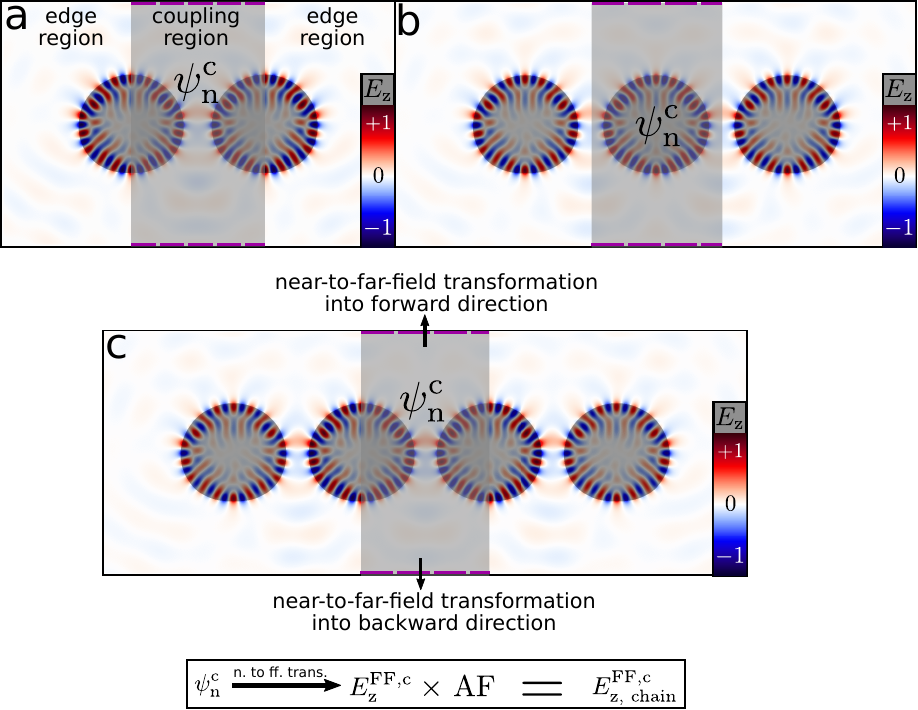}
	\caption{\textbf{a - c:} Field distributions $E_\text{z}$ of 2 ($d/R=0.58$), 3 ($d/R=0.52$) and 4 ($d/R=0.46$) resonators, respectively. The field $\psi_n^\text{c}$ inside the gray region containing the coupling information is used to compute the far fields $E_{\text{z}}^{\text{FF,c}}$ into the forward and backward direction. 
	}
	\label{fig:S3}
\end{figure}

\noindent The generalized array factor model takes the coupling of the resonators into account.
In the strong coupling regime, $\psi_n$ and $E_{\text{z,array}}^{\text{FF}}$ of the bare single cavity are replaced by the respective quantities $\psi_n^\text{c}$ and $E_{\text{z,array}}^{\text{FF,c}}$ of the single coupled cavity as illustrated in FIG.~\ref{fig:S3} a - c.\\
%
The far field of the resonator array in the strong coupling regime is modeled by multiplying the far fields of the single coupled cavity with the array factor:

\begin{equation}
E_{\text{z,array}}^{\text{FF,c}} (\theta) \propto E_{\text{z}}^{\text{FF,c}}(\theta) \sum_{n=1}^{N}\alpha_{n}e^{-ikD(n-1)\sin{\theta}}.
\end{equation}   

\noindent For reasons of simplicity, the coefficients $\alpha_n$ are set to unity. The far fields have to be computed within a small $\theta$-range ($\Delta\theta=\pm20^\circ$), separately for the forward and backward direction since the $\psi_n^\text{c}$-region does not enclose the total FDTD-cell.

\vspace{1cm}
\section{Directionality reversion depending on number of array elements  }
\label{sec:4}

\begin{figure}[h]
	\centering
	\includegraphics[width=16cm]{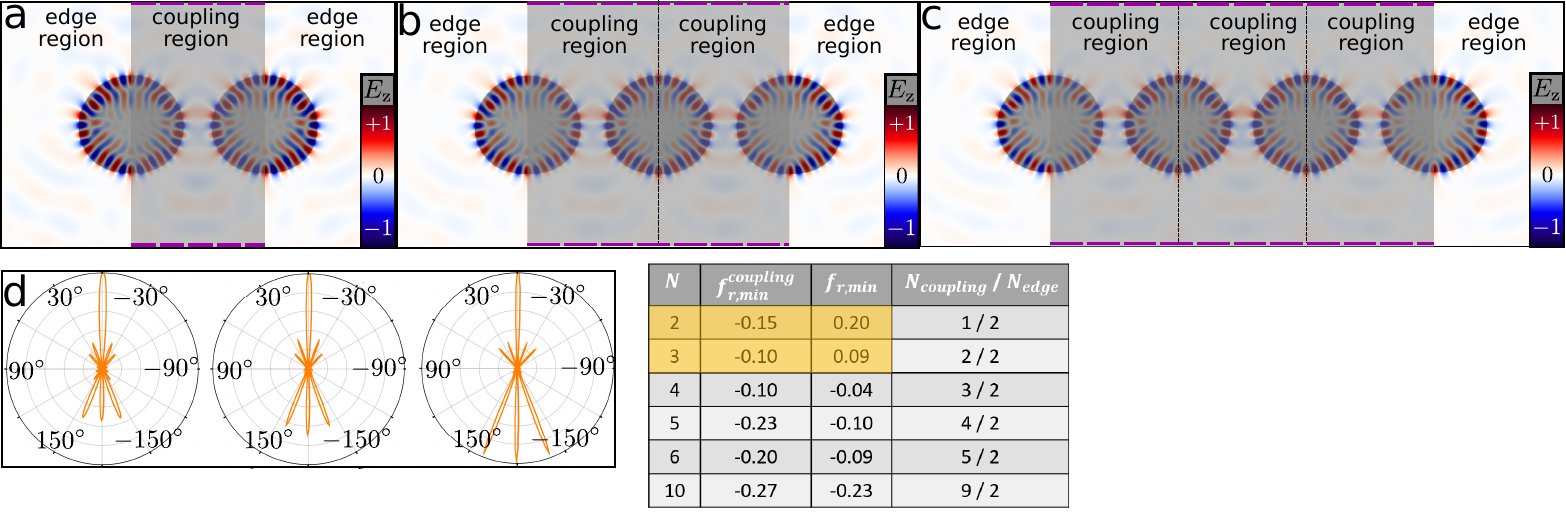}
	\caption{\textbf{a - c:} Field distributions $E_\text{z}$ of 2 ($d/R=0.52$), 3 ($d/R=0.46$) and 4 ($d/R=0.40$) resonators, respectively. $d/R$-values correspond to the minima of $f_\text{r}$,  respectively. \textbf{d:} Corresponding far-field intensities of the array modes shown in figures a - c. \textbf{Table:} Comparison of the emission ratio values computed from the coupling regions $f_{\text{r,min}}^{\text{couping}}$ and of those computed from the entire FDTD-cell $f_{\text{r,min}}$  (including coupling and edge regions, as indicated in figures a - c ) for various resonator numbers $N$.   
	}
	\label{fig:S4}
\end{figure}

\noindent The feature of directionality reversion is already present for a 2 and 3 resonator array, as can be concluded from the emission ratio values highlighted in the table of FIG.~\ref{fig:S4}. The emission ratio values $f_{\text{r,min}}^{\text{couping}}$ are computed from the fields inside the coupling regions. Actually, the $f_{\text{r,min}}^{\text{couping}}$-values for 2 and 3 resonators are negative which indicates that more energy is emitted backwards than forwards inside the coupling regions. The fact that the total (including coupling and edge regions) emission ratio $f_{\text{r,min}}$ for 2 and 3 resonators is positive suggests that the emission from the edge regions is too dominant and masks the directionality reversion. 
We expect that with increasing number of array elements $N$ the emission from the edge regions should be less relevant.  For $N\geq4$ both $f_{\text{r,min}}^{\text{couping}}$ and $f_{\text{r,min}}$ are negative which indicates that the directionality reversion of the entire array occurs. Furthermore, it indicates that the emission from the edge regions is less relevant. Interestingly, the directionality reversion occurs if the number of coupling regions exceeds the number of edge regions, as displayed in the right column of the table in FIG.~\ref{fig:S4}.

\section{Resonance dependence of directionality reversion and super-directionality}
\label{sec:5}

\begin{figure}[h]
	\centering
	\includegraphics[width=16cm]{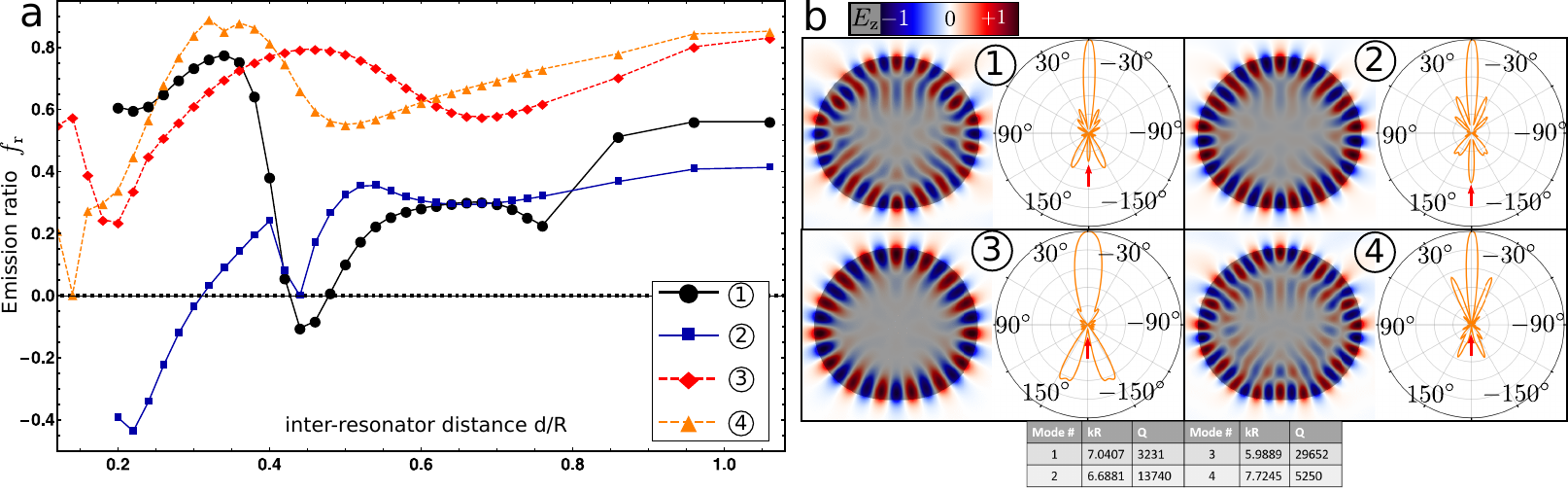}
	\caption{\textbf{a:} Emission ratio of a 5-resonator array excited at for four different modes. \textbf{b:} collocation of the corresponding uncoupled (single) resonator modes and far fields.
	}
	\label{fig:S5}
\end{figure}

\noindent Since far fields are a resonance dependent feature, we expect the emission ratio to depend strongly on the chosen resonance. As a consequence, the directionality reversion and super-directionality will depend strongly on the resonance as our findings in FIG.~\ref{fig:S5} illustrate. 
FIG.~\ref{fig:S5} a shows the emission ratio for four different modes. The corresponding modes and far fields of the single resonator are displayed in FIG.~\ref{fig:S5} b. All four modes show similar far-field features, namely a pronounced main lobe into the forward direction ($\theta=0^\circ$) and  two side lobes into the directions with $\theta=\pm150^\circ$. The crucial difference between the modes 1,2 (showing directionality reversion) and the modes 3,4 (without or weak directionality reversion) is the emission into the backwards direction $\theta=\pm180^\circ$ as indicated by red arrows in the far-field plots in FIG.~\ref{fig:S5} b. \\ 
This suggests that the backward-emission peak of the uncoupled resonator plays an important role for the inter-resonator coupling (work in progress).\\

